\documentclass[aps,jctc,reprint,superscriptaddress]{revtex4-1}

\usepackage{amsmath}
\usepackage{graphicx}
\usepackage{color}
\usepackage{hyperref}
\usepackage{enumitem}

\newcommand{\gj}[1]{\textcolor{black}{#1}}
\newcommand{\fs}[1]{\textcolor{black}{#1}}

\begin{document}

\title{Frequency-dependent Hydrodynamic Interaction Between Two Solid Spheres}

\author{Gerhard Jung}
\email{jungge@uni-mainz.de}
\affiliation{Institut f\"ur Physik, Johannes Gutenberg-Universit\"at Mainz, 
Staudingerweg 9, 55128 Mainz, Germany}
\affiliation{Graduate School of Excellence Materials Science in Mainz, Staudingerweg 9, 55128 Mainz, Germany}
\author{Friederike Schmid}
\email{friederike.schmid@uni-mainz.de}
\affiliation{Institut f\"ur Physik, Johannes Gutenberg-Universit\"at Mainz, 
Staudingerweg 9, 55128 Mainz, Germany}

\begin{abstract}

Hydrodynamic interactions play an important role in many areas of soft matter science. In simulations with implicit solvent, various techniques such as Brownian or Stokesian dynamics explicitly include hydrodynamic interactions \emph{a posteriori} by using hydrodynamic diffusion tensors derived from the Stokes equation. However, this equation assumes the interaction to be instantaneous which is an idealized approximation and only valid on long time scales. In the present paper, we go one step further and analyze the time-dependence of hydrodynamic interactions \fs{between finite-sized particles} in a compressible fluid on the basis of the linearized Navier-Stokes equation. The theoretical results show that \gj{at high frequencies} the compressibility of the fluid has a significant impact on the frequency-dependent pair interactions.

The predictions of hydrodynamic theory are compared to molecular dynamics simulations of two nanocolloids in a Lennard-Jones fluid. For this system we reconstruct memory functions by extending the inverse Volterra technique. The simulation data agree very well with the theory, therefore, the theory can be used to implement dynamically consistent hydrodynamic interactions in the increasingly popular field of non-Markovian modeling.

\end{abstract}

\maketitle


\section{Introduction}

It is well known that hydrodynamic interactions between different macromolecules crucially influence many physical processes (some examples: colloid diffusion with applications to microrheology \cite{Mason1995,Cordoba2012a}, colloid crystallization \cite{Vermant2005}, protein folding and diffusion \cite{FrembgenKesner2009} or polymer aggregation \cite{Tomilov2013}). In coarse-graining, however, this hydrodynamic interaction is often lost because of the use of implicit solvent potentials. Therefore, many mesoscopic models explicitly include hydrodynamic interactions between molecules to achieve an accurate representation of the underlying fine-grained system. The most important simulation techniques that explicitly account for hydrodynamic interactions are Brownian dynamics \cite{Ermak1975,VanGunsteren1982} (including fluctuations) and Stokesian dynamics  \cite{Brady1988} (without fluctuations). 

The input for these techniques is the hydrodynamic diffusion tensor, describing the hydrodynamic interactions between two spherical particles on a pairwise level. This diffusion tensor can be determined by the method of reflections \cite{Happel1963,Dhont1996}.  In first order it was derived by Oseen (Oseen tensor) \cite{Oseen1911} and in second order by Rotne and Prager (Rotne-Prager tensor) \cite{Rotne1969}. The velocity fields used as input for the method of reflections are solutions of the steady-state Stokes equations \gj{\cite{Stokes1845}},
\begin{eqnarray}
\eta \nabla^2 \mathbf{u}(\mathbf{r}) - \nabla p(\mathbf{r}) + \mathbf{F}_1^{(1)}(\mathbf{r}) &=& 0 \quad\\
\nabla \cdot \mathbf{u}(\mathbf{r}) &=& 0
\end{eqnarray}
which describe the velocity-field $ \mathbf{u}(\mathbf{r}) $ that is created by a force \gj{density} $ \mathbf{F}_1^{(1)}(\mathbf{r}) $ acting on particle 1 in an incompressible fluid. The parameter $ \eta $ describes the shear viscosity of the underlying fluid and $ p(\mathbf{r}) $ the pressure field.

The steady-state description is limited to cases, where the relaxation of the fluid can be assumed to be much faster than the relaxation of the macromolecules. In the case of overdamped dynamics (Brownian or Stokesian dynamics), this assumption is reasonable. On small time scales, however, one needs to consider frequency-dependent interactions by applying the unsteady Stokes equations \gj{\cite{Stokes1845}},
\begin{eqnarray}
\rho \frac{\partial \mathbf{u}(\mathbf{r},t)}{\partial t} + \eta \nabla^2 \mathbf{u}(\mathbf{r},t) - \nabla p(\mathbf{r},t) + \mathbf{F}_1^{(1)}(\mathbf{r},t) &=& 0 \quad\label{eq:USE}\\
\nabla \cdot \mathbf{u}(\mathbf{r},t) &=& 0 \quad
\end{eqnarray}
with constant fluid mass density $ \rho $. The most prominent consequence of adding the time-derivative in Eq.~(\ref{eq:USE}) is that the motion of the macromolecule will create vortexlike structures propagating with finite velocity. This is the origin of the famous Basset force, which leads to the long-time tail in the velocity auto-correlation function of a macromolecule submerged in a fluid \cite{Basset1888,Alder1970,Liverpool2005}. The unsteady Stokes equation can also be used to calculate cross-correlations between two macromolecules. For that case Ardekani \emph{et al.}~\cite{Ardekani2006} derived a frequency-dependent generalization of the Oseen tensor.

The Stokes equations are, however, restricted to incompressible fluids. This \gj{implies} that the speed of sound is infinite. In the present paper we go one step further and consider the linearized Navier-Stokes equations \cite{Mazur1974},
\begin{eqnarray}
\rho_e \frac{\partial \mathbf{u}(\mathbf{r},t)}{\partial t} &=& - \nabla \mathbf{P}(\mathbf{r},t) + \mathbf{F}_1^{(1)}(\mathbf{r},t)  \nonumber\\
\frac{\partial \rho(\mathbf{r},t)}{\partial t} &=&  -\rho_e \nabla \cdot \mathbf{u}(\mathbf{r},t)
\label{eq:lns}
\end{eqnarray}
with the uniform equilibrium density of the fluid $ \rho_e $. The pressure tensor $ P_{ij}(\mathbf{r},t) $ is given by,
\begin{equation}
P_{ij} = p \delta_{ij} -\eta (\partial_i u_j + \partial_j u_i) + \left (\frac{2}{3} \eta - \zeta \right ) \nabla \cdot \mathbf{u} \delta_{ij}
\label{eq:pressure_tensor},
\end{equation}
with bulk viscosity $ \zeta $. For these equations we will derive similar relations as presented by Ardekani \emph{et al.} \cite{Ardekani2006} generalized to compressible fluids. \fs{We can show, that the resulting equations are equivalent to a comparable generalization obtained by C\'{o}rdoba \emph{et al.} \cite{Cordoba2013}. Subsequently, we derive a correction for finite-sized particles, which enables the quantitative comparison to results from molecular dynamics (MD) simulations.}

\fs{To this end,} we perform MD simulations of two nanocolloids in a Lennard-Jones fluid, each trapped in a harmonic potential. For this system we determine self- and pair-memory functions by extending the inverse Volterra method \cite{Shin2010} as well as velocity correlation functions. \gj{Compared to previous studies of single particles in fluids \cite{Kohale2010,Grimm2011,Theers2016}, the main contribution of this work is to move from the analysis of self-correlations to cross-correlations of a pair of particles. This finally allows to better understand the frequency-dependence of particle pair interactions.} With this paper we therefore combine recent advancements in hydrodynamic theory \cite{Ardekani2006,Vazquez-Quesada2013,Cordoba2013,Wang2017}, \gj{computer simulations and experiments of single particles in (viscoelastic) fluids \cite{Kohale2010,Grimm2011,Theers2016}} and non-Markovian modeling \cite{Li2017,Jung2017}.

Our paper is organized as follows: In Sec.~\ref{sec:theory}, we will explain the theoretical details and derive the frequency-dependent interaction between two solid spheres. To compare the theory to molecular dynamics (MD) simulations  of nanocolloids in a Lennard-Jones fluid, we characterize the fluid in Sec. \ref{sec:fluid}. This enables us to determine all necessary input parameters like viscosities, hydrodynamic radius and speed of sound. In Sec. \ref{sec:memory}, we will shortly recapitulate the notion of memory and the generalized Langevin equation. We then can establish relations for time correlation functions in Fourier space and show how to reconstruct the pair memory function from MD time correlation functions. These results will be compared to theory in Sec. \ref{sec:results} at the example of two nanocolloids submerged in a Lennard-Jones fluid. We summarize and conclude in Sec. \ref{sec:conclusion}.


\section{Theory}
\label{sec:theory}

The starting point of the derivation is the set of linearized Navier-Stokes equations (see Eqs.~(\ref{eq:lns}) and (\ref{eq:pressure_tensor})) combined with a linearized relation between pressure and density gradient,
\begin{equation}
\nabla p(\mathbf{r},t) = c_0^2 \nabla \rho(\mathbf{r},t) ,
\end{equation}
with the adiabatic speed of sound $ c_0 $ of the fluid. To solve these equations we transform them into Fourier space, which leads to the following coupled differential equations:
\begin{eqnarray}
(-{\rm i} \omega \rho_e - \eta \Delta) \mathbf{\hat{u}}(\mathbf{r,\omega}) &=& - \mu \nabla \hat{\rho} (\mathbf{r,\omega}) + \mathbf{\hat{F}}_1^{(1)}(\mathbf{r,\omega})\\
(- \omega^2 - c^2 \Delta) \hat{\rho}(\mathbf{r,\omega}) &=& - \nabla \cdot \mathbf{\hat{F}}_1^{(1)}(\mathbf{r,\omega})  ,
\end{eqnarray}
with the Fourier transform of the velocity field
\begin{equation}
\mathbf{\hat{u}}(\mathbf{r,\omega}) = \int_{-\infty}^{\infty} e^{{\rm i}\omega t} \mathbf{u}(\mathbf{r,t})
\end{equation}
and similarly the Fourier transforms of the density field $ \hat{\rho}(\mathbf{r,\omega}) $ and the external force \gj{density} $ \mathbf{\hat{F}}_1^{(1)}(\mathbf{r,\omega}) $. The coefficient $ \mu $ and the frequency-dependent speed of sound $ c $ are given by
\begin{eqnarray}
\mu &=& c_0^2 - {\rm i}\omega \left (\frac{1}{3} \eta + \zeta \right ) \rho_e^{-1}  \\
c^2 &=& c_0^2 - {\rm i}\omega \left (\frac{4}{3} \eta + \zeta \right ) \rho_e^{-1},\text{ with Im}(c) > 0.
\end{eqnarray}
We will now follow the derivation of Bedeaux \emph{et al.} (see Eqs.~(2.16)-(2.22) in \cite{Mazur1974}) and introduced the transversal and longitudinal Green's functions $ G_\text{tr} $ and $ G_\text{l} $,
\begin{eqnarray}
G_{tr}(\mathbf{r,\omega}) &=& (4 \pi \eta r)^{-1} \exp(-\alpha r)\\
G_{l}(\mathbf{r,\omega}) &=& (4 \pi c^2 r)^{-1} \exp(-{\rm i} \omega r/c)  ,
\end{eqnarray}
with distance $ r = \left|  \mathbf{r}\right|  $ and $ \alpha = (-{\rm i} \omega \rho_e/ \eta)^{\frac{1}{2}} $ with $ \text{Re}(\alpha)>0 $. This allows us to write down a formal solution of the linearized Navier-Stokes equation:
\begin{eqnarray}
\mathbf{\hat{u}}(\mathbf{r,\omega}) &=& \int \text{d}\mathbf{r'}  ( G_{tr}(\mathbf{r-r',\omega}) + \alpha^{-2} \frac{\partial}{\partial \mathbf{r'}} \otimes \frac{\partial}{\partial \mathbf{r'}} \label{eq:velocityfield} \\ & \times & \left[ \eta^{-1} c^2 G_{l}(\mathbf{r-r',\omega}) -G_{tr}(\mathbf{r-r',\omega}) \right] ) \cdot \mathbf{\hat{F}}_1^{(1)}(\mathbf{r',\omega}) \nonumber .
\end{eqnarray}

\subsection{Point Force}
\label{sec:theoryPF}

With the formal solution for the velocity field (see Eq.~(\ref{eq:velocityfield})) and the assumption of a point force acting on the center of particle 1 in the origin,
\begin{equation}
\mathbf{\hat{F}}_1^{(1)}(\mathbf{r,\omega}) = \mathbf{\hat{F}}_1^{(1)}(\omega)\delta(\mathbf{r}),
\end{equation}
we can write down the velocity field explicitly
\begin{equation}
\mathbf{\hat{u}}(\mathbf{r,\omega}) = - \frac{1}{8 \pi \eta} \left[ A(\mathbf{r,\omega}) \mathbf{\hat{F}}_1^{(1)}(\omega) + B(\mathbf{r,\omega}) (\mathbf{\hat{F}}_1^{(1)}(\omega) \cdot \mathbf{n}) \mathbf{n} \right]
\label{eq:velocityfieldPF}
\end{equation}
with $ \mathbf{n} = \mathbf{r}/r $ and the parameters $ A(\mathbf{r,\omega}) $ and $ B(\mathbf{r,\omega}) $ given by:

\begin{eqnarray}
A(\mathbf{r,\omega}) &=& A_\text{l}(\mathbf{r,\omega}) + A_\text{tr}(\mathbf{r,\omega})  \\
B(\mathbf{r,\omega}) &=& B_\text{l}(\mathbf{r,\omega}) + B_\text{tr}(\mathbf{r,\omega}),
\end{eqnarray}
with
\begin{eqnarray}
A_\text{l}(\mathbf{r,\omega}) &=& - \left( \frac{2 {\rm i} \omega}{r^2\alpha^2 c} + \frac{2}{r^3\alpha^2} \right) e^{-{\rm i} \omega r/c} \nonumber\\
A_\text{tr}(\mathbf{r,\omega}) &=& \left( \frac{2}{r} + \frac{2}{\alpha r^2} + \frac{2}{\alpha^2r^3} \right) e^{-\alpha r} \nonumber\\
B_\text{l}(\mathbf{r,\omega}) &=& \left( - \frac{\omega^2}{c^2}\frac{2}{\alpha^2 r} + \frac{6 {\rm i} \omega}{r^2\alpha^2 c} + \frac{6}{r^3\alpha^2} \right) e^{-{\rm i} \omega r/c} \nonumber\\
B_\text{tr}(\mathbf{r,\omega}) &=& - \left( \frac{2}{r} + \frac{6}{\alpha r^2} + \frac{6}{\alpha^2r^3} \right) e^{-\alpha r}.
\label{eq:ab_parameter}
\end{eqnarray}

Since the governing equations are linear, it is possible to decompose the velocity field and apply the method of reflections \cite{Happel1963,Dhont1996}. This means that we evaluate the velocity field $ \mathbf{\hat{u}}(\mathbf{r,\omega}) $ at the center of particle 2 and determine the friction force $ \mathbf{\hat{F}}_2^{(1)}(\omega) = - \hat{\gamma}(\omega) ( \mathbf{\hat{v}}_2 - \mathbf{\hat{u}}(\mathbf{r,\omega}) ) $, where $ \mathbf{\hat{v}}_2 $ is the velocity of particle 2 and $ \hat{\gamma}(\omega) $ the frequency-dependent response  of the \emph{solitary} sphere in the flow $ \mathbf{\hat{u}}(\omega) $. \gj{An explicit expression for $ \hat{\gamma}(\omega) $ will be given at the end of Sec.~\ref{sec:theoryUV}}. \gj{Here, we also assumed a uniform flow and disregarded Faxen's theorem. This approximation is therefore only valid at small radius to distance ratios $ R/d $.} For details we refer to Ref.~\cite{Ardekani2006} Eqs.~(28)-(34) as they use the same formalism, just with different parameters $ A(\mathbf{r,\omega}) $ and $ B(\mathbf{r,\omega}) $.

The final result can be written as,
\begin{equation}
\hat{F}_{1,\parallel}(d,\omega) = -\hat{\gamma}_{11,\parallel}(d,\omega) \hat{v}_{1,\parallel}(\omega) - \hat{\gamma}_{12,\parallel}(d,\omega) \hat{v}_{2,\parallel}(\omega),
\label{eq:hydrotheory_first}
\end{equation}
with the frequency-dependent self- and cross-memory functions,
\begin{eqnarray}
\hat{\gamma}_{11,\parallel}(d,\omega) &=& \frac{\hat{\gamma}(\omega)}{1-D_\parallel(d,\omega)^2}\label{eq:memory_theory0}\\
\hat{\gamma}_{12,\parallel}(d,\omega) &=& - \frac{\hat{\gamma}(\omega)D_\parallel(d,\omega)}{1-D_\parallel(d,\omega)^2},
\label{eq:memory_theory}
\end{eqnarray}
and the Fourier transformed velocities of particle $ i $, $ \hat{v}_{i,\parallel}(\omega) $.
Here, $ \parallel $ denotes the direction parallel to the line of centers of the two spheres. The same equations hold for the perpendicular direction $ \perp $. The coefficients $ D_\parallel^{\text{PF}}(d,\omega) $ and $ D_\perp^{\text{PF}}(d,\omega) $ for the point force are then given by:
\begin{eqnarray}
D_\parallel^{\text{PF}}(d,\omega) &=& \frac{\hat{\gamma}(\omega)}{8 \pi \eta} \left (A(d,\omega)+B(d,\omega) \right ) \label{eq:DparPF}\\
D_\perp^{\text{PF}}(d,\omega) &=& \frac{\hat{\gamma}(\omega)}{8 \pi \eta} A(d,\omega),\label{eq:DperpPF}
\end{eqnarray}
where $ \mathbf{d} $ is the distance vector between the two spheres and $ d = \left| \mathbf{d}\right|  $. \fs{Our solution is equivalent to Eq.~(7) in Ref.~\cite{Cordoba2013}, if one uses the mapping $ G^*=-i\omega \eta $ and $ 1/3G^*+K^*=-i\omega(1/3\eta +\zeta) + \rho_e c_0^2 $.}

These equations represent the frequency-dependent response of two solid spheres - under the assumption that the velocity fields are produced by point forces. \gj{Similar to the assumption of uniform flow, this restricts the applicability of the relations to small ratios $ R/d $.} Analogous to Ardekani \emph{et al.} \cite{Ardekani2006} we now improve this Ansatz by considering an unsteady velocity field created by the movement of a \fs{finite-sized particle}.

\subsection{Unsteady velocity field} 
\label{sec:theoryUV}

The Fourier transformed velocity field described in Eq.~(\ref{eq:velocityfieldPF}) can be written as,
\begin{eqnarray}
\mathbf{\hat{u}}(\mathbf{r,\omega}) = &-&\mathbf{\hat{e}}_\mathbf{F} \Delta \hat{\Psi}_\text{tr}(\mathbf{r,\omega}) \nonumber \\&+& (\mathbf{\hat{e}}_\mathbf{F} \cdot \nabla) \nabla \left (\hat{\Psi}_\text{l}(\mathbf{r,\omega}) + \hat{\Psi}_\text{tr}(\mathbf{r,\omega})\right ),
\label{eq:velocityfield_ansatz}
\end{eqnarray}
with $ \hat{\Psi}_l(\mathbf{r,\omega}) = \frac{Q_\text{l}}{r} e^{-{\rm i} \omega r/c}, $  $ \hat{\Psi}_\text{tr}(\mathbf{r,\omega})= \frac{Q_\text{tr}}{r} e^{-\alpha r}  $ and $ \mathbf{\hat{e}}_\mathbf{F} = \mathbf{\hat{F}}_1^{(1)} /  \hat{F}_1^{(1)} $. By choosing 
\begin{eqnarray}
Q_\text{l} &=& -\frac{2}{\alpha^2} \frac{1}{8 \pi \eta} \hat{F}_1^{(1)}\\
Q_\text{tr} &=& \frac{2}{\alpha^2} \frac{1}{8 \pi \eta} \hat{F}_1^{(1)},
\end{eqnarray}
we recover Eq.~(\ref{eq:velocityfieldPF}) and hence Eqs.~(\ref{eq:DparPF}) and (\ref{eq:DperpPF}).

To generalize this solution for spheres with finite radius, we use Eq.~\ref{eq:velocityfield_ansatz} as an Ansatz with adjustable parameters $ Q_\text{l} $, $ Q_\text{tr} $, which are chosen such that the flow field satisfies the appropriate boundary conditions at the surface of the spheres. This Ansatz is similar to the  Burgers' solution in Ref.~\cite{Burgers1938} with the extension that the longitudinal motion is not instantaneous due to the compressibility of the fluid (for Burgers' solution the limit $ c \rightarrow \infty $ has to be taken). Here, we consider no-slip boundary conditions, i.e.
\begin{eqnarray}
 \mathbf{\hat{u}}(\mathbf{r},\omega) &=& - \mathbf{\hat{F}}_1^{(1)}/\hat{\gamma}(\omega) \quad \text{at } \left| \mathbf{r}\right| =R \label{eq:boundary_uvf}\\
\mathbf{\hat{u}}(\mathbf{r},\omega) &=& 0 \hspace*{1.95cm} \text{as } \left| \mathbf{r}\right| \rightarrow \infty ,
\end{eqnarray}
with the radius $ R $ of the solid sphere. From these boundary conditions we can determine the coefficients $ Q_\text{l} $, $ Q_\text{tr} $:
\begin{eqnarray}
Q_\text{l} &=& \frac{{\hat{F}}_1^{(1)}}{\hat{\gamma}(\omega)} \frac{c^2 R (3+3R\alpha+R^2\alpha^2) e^{{\rm i} \omega R/c}}{w^2(1+\alpha R  + \alpha^2 R^2) - 2 \alpha^2 c^2 - 2 {\rm i} \alpha^2cRw} \nonumber\\
&\equiv& - \frac{{\hat{F}}_1^{(1)}}{\hat{\gamma}(\omega)} \hat{Q}_\text{l}, \\
Q_\text{tr} &=&-\frac{{\hat{F}}_1^{(1)}}{\hat{\gamma}(\omega)} \frac{R (3c^2+3R{\rm i}c\omega-R^2\omega^2) e^{R \alpha}}{w^2(1+\alpha R  + \alpha^2 R^2) - 2 \alpha^2 c^2 - 2 {\rm i} \alpha^2cRw}\nonumber\\
&\equiv& - \frac{{\hat{F}}_1^{(1)}}{\hat{\gamma}(\omega)} \hat{Q}_\text{tr}.
\end{eqnarray}
Similar to Sec. \ref{sec:theoryPF} we can apply the method of reflections to determine the final solution for the self- and cross-memory functions. In this way we can determine the coefficients $ D_\parallel^{\text{UV}}(d,\omega) $ and $ D_\perp^{\text{UV}}(d,\omega) $ for the unsteady velocity field:
\begin{eqnarray}
D_\parallel^{\text{UV}}(d,\omega) &=& D_{\parallel,l}^{\text{UV}}(d,\omega) +  D_{\parallel,tr}^{\text{UV}}(d,\omega) \label{eq:uvf_result_par}\\
D_\perp^{\text{UV}}(d,\omega) &=& D_{\perp,l}^{\text{UV}}(d,\omega) +  D_{\perp,tr}^{\text{UV}}(d,\omega),
\end{eqnarray}
with
\begin{eqnarray}
D_{\parallel,\text{l}}^{\text{UV}}(d,\omega) &=& \hat{Q}_\text{l} \left ( \frac{2}{d^3} + \frac{2{\rm i} \omega}{c d^2} - \frac{\omega^2}{c^2 d} \right ) e^{-{\rm i} \omega d/c}\nonumber\\
D_{\parallel,\text{tr}}^{\text{UV}}(d,\omega) &=& 2 \hat{Q}_\text{tr} \left ( \frac{1}{d^3} + \frac{\alpha}{d^2}  \right ) e^{-\alpha d} \nonumber\\
D_{\perp,\text{l}}^{\text{UV}}(d,\omega) &=& - \hat{Q}_\text{l} \left ( \frac{1}{d^3} + \frac{{\rm i} \omega}{c d^2} \right ) e^{-{\rm i} \omega d/c}\nonumber\\
D_{\perp,\text{tr}}^{\text{UV}}(d,\omega) &=& - \hat{Q}_\text{tr} \left ( \frac{1}{d^3} + \frac{\alpha}{d^2} + \frac{\alpha^2}{d} \right ) e^{-\alpha d}. \label{eq:uvf_result_par_l}
\end{eqnarray}
To close Eqs.~(\ref{eq:memory_theory0}) and (\ref{eq:memory_theory}) we need to find an explicit expression for the frequency-dependent response $ \hat{\gamma}(\omega) $ of the solitary sphere. In the following we will use the solution by Bedeaux \emph{et al.} \cite{Mazur1974} (rewritten into the form of Ref.~\cite{Metiu1977}) for a compressible fluid with no-slip boundary conditions:
\begin{equation}
\hat{\gamma}(\omega) = \frac{4 \pi}{3} \eta R X^2 \left[ (1-Y)Q+2(X-1)P\right],
\label{eq:frequency-dependent-reponse-single}
\end{equation}
with
\begin{eqnarray}
X&=&-\alpha R\\
Y&=&-{\rm i}\omega R/c\\
Q&= &\frac{3}{\Delta} \left ( 3- 3X +X^2 \right)\\
P&=&-\frac{3}{\Delta} (Y^2-3Y+3)\\
\Delta&=& 2X^2(3-3Y+Y^2) -Y^2(3-3X+X^2) .
\end{eqnarray}
\begin{figure}
\includegraphics[scale=1]{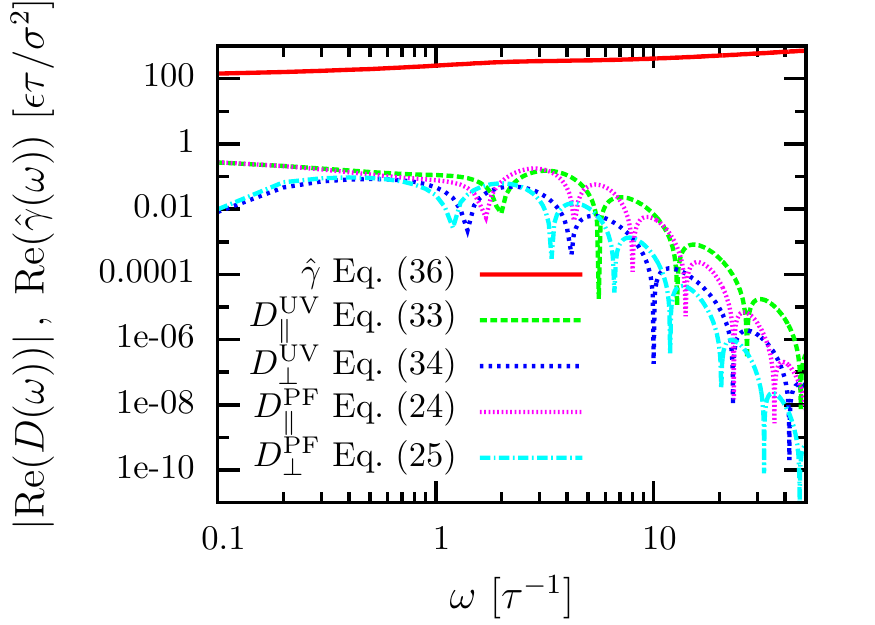}
\caption{Frequency-dependence of the response functions $ \hat{\gamma}(\omega) $ and $ D(\omega) $ as derived in Sec.~\ref{sec:theory}. The input parameters are obtained in Sec.~\ref{sec:fluid}, the distance was chosen to be $ d = 8.5\,\sigma $. }
\label{fig:frequency_dependence}
\end{figure}
In Fig.~\ref{fig:frequency_dependence} the frequency-dependence of the response functions that were derived in this section are illustrated. The response function of the solitary sphere $ \hat{\gamma}(\omega) $ is divergent for $ \omega \rightarrow \infty $, leading to a divergent instantaneous response $ \gamma(t=0) $. We will therefore expect discrepancies between theory and simulations for high frequencies or small times (as reported in Ref.~\cite{Theers2016}, Fig.~1). This can be explained by the particle character of the fluid for high frequencies, that is not captured in the continuum description of hydrodynamic theory. The multiplication of $ \hat{\gamma}(\omega) $ with $ D_{\parallel/\perp}(\omega) $ to determine the cross-memory function $ \gamma_{12,\parallel}(\omega) $, however, corresponds to a low-pass filter (see Eq.~(\ref{eq:memory_theory}) and Fig.~\ref{fig:frequency_dependence}). Thus high frequency contributions to the cross-memory function are damped.


\section{Characterization of the fluid}
\label{sec:fluid}

In the simulations, we considered the diffusion of a pair of nanocolloids in a Lennard-Jones (LJ) fluid. The LJ particles are initially placed on a fcc-lattice with lattice constant $ a = 1.71 \sigma $ and therefore have a reduced density of $ \rho^* = \rho \sigma^3 = 0.8 $. The reduced temperature was set to $ T^* = k_\text{B} T / \epsilon = 1.0  $. The LJ diameter $ \sigma $, energy $ \epsilon $ and time $ \tau = \sigma \sqrt{m / \epsilon} = 1 $ are defining the length, energy and time units of the simulation. The LJ cutoff was set to $ r_\text{c} = 2.5\, \sigma $ and the particle mass to $ m^*= 1\,m $. The nanocolloids were created by fixing the inter-particle distances of 80 LJ particles in the fcc-lattice with a radius $ R = 3\, \sigma $. \gj{The resulting nanocolloids are not perfectly isotropic} (see Fig.~\ref{fig:rdf}), however, the discrepancies do not have a significant impact on the interaction. The nanocolloids are trapped in a harmonic potential, so that the distance between the nanocolloids is approximately constant. We applied periodic boundary conditions in all three dimensions to the cubic simulation box and equilibrated the system using a Langevin thermostat. The system was then integrated with a time step of $ \Delta t_\text{MD} = 0.001\,\tau $ in the \emph{NVE}-ensemble. The simulations were performed with the simulation package \emph{Lammps} \cite{Plimpton1995}.

To calculate theoretical predictions for the memory functions using Eq.~(\ref{eq:memory_theory}), it is necessary to determine the input parameter that characterize the fluid, namely the shear viscosity $ \eta $, bulk viscosity $ \zeta $, hydrodynamic radius $ R_H $ and speed of sound $ c_0 $.

\subsection{Shear and bulk viscosity}

To determine the shear viscosity $ \eta $ and bulk viscosity $ \zeta $ we simulated a bulk LJ fluid without colloids and box size $ L_B = 43.09\,\sigma $. The viscosities can  then by calculated using the Green-Kubo relations \cite{Green1954,Kubo1957, Jung2016}:
\begin{eqnarray}
\eta &=& \frac{V}{k_B T} \int_{0}^{\infty} \text{d}t' \left\langle \sigma_{xy}(t') \sigma_{xy}(0) \right\rangle, \label{eq:shear_viscosity} \\
\zeta &=& \frac{V}{k_B T} \int_{0}^{\infty} \text{d}t' \left\langle \delta p(t') \delta p(0) \right\rangle,
\end{eqnarray}
with the off-diagonal component of the stress tensor $ \sigma_{xy}(t) $ and the pressure fluctuations $ \delta p(t) = \frac{1}{3} \sum_{\alpha} \sigma_{\alpha \alpha}(t) - \left\langle \sigma_{\alpha \alpha} \right\rangle $. 

To validate the results for the shear viscosity, we have also performed non-equilibrium molecular dynamics simulations (NEMD) using the M\"uller-Plathe method \cite{Muller-Plathe1999}. In this method, the simulation box is divided into several slabs. \emph{The momentum} of particles in different slabs is frequently exchanged to create a shear flow. This mechanism conserves total momentum and energy and allows the determination of the shear viscosity $ \eta $ by determining the momentum flux between the slabs (for details, see Ref.~\cite{Muller-Plathe1999}). The results can be found in Tab.~\ref{tab:viscosity}. The agreement between Green-Kubo relations and NEMD is very good, therefore, the values for the viscosities are reliable. 

\begin{table}
\begin{tabular}{|c|c|c|}
\hline  & GK & NEMD  \\ 
\hline shear viscosity $ \eta $ $ \gj{[\epsilon \tau/\sigma^3]} $ & $  2.11 \pm 0.01 $ & $  2.11 \pm 0.03 $   \\ 
\hline bulk viscosity $ \zeta $ $ \gj{[\epsilon \tau/\sigma^3]} $ & $  0.88 \pm 0.01 $ & -  \\ 
\hline 
\end{tabular} 
\caption{Shear and bulk viscosity of a Lennard-Jones fluid with $ T^* = 1 $ and $ \rho^* = 0.8 $. The values are obtained using Green-Kubo relations (GK) and non-equilibrium molecular dynamics simulations (NEMD).}
\label{tab:viscosity}
\end{table}

\subsection{Hydrodynamic radius}

\begin{figure}
\includegraphics[scale=1]{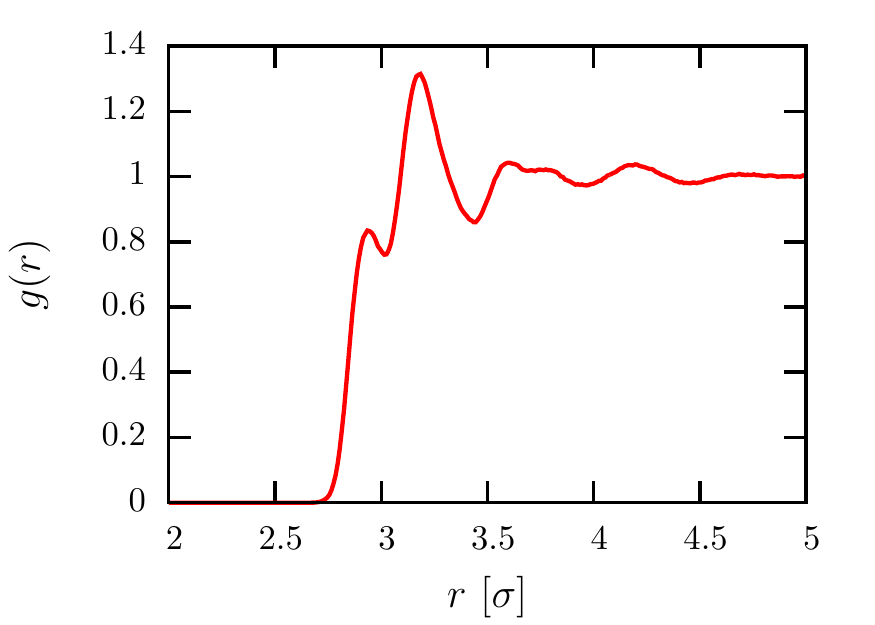}
\caption{Radial distribution function $ g(r) $ between the \gj{center-of-mass} of the nanocolloid  and the solvent particles. The first solvation shell is located at a distance of around $ r_m = 3.2\,\sigma $. \gj{The small peak at around $ 2.9\,\sigma  $ occurs due to the structure of the nanocolloid, which is built from LJ particles (see Sec.~\ref{sec:fluid})}. }
\label{fig:rdf}
\end{figure}

The most straightforward way to determine the hydrodynamic radius $ R_H $ is to calculate the radial distribution function $ g(r) $ (RDF, see Fig.~\ref{fig:rdf}). The first maximum $ r_m $ in the RDF corresponds to the distance between the nanocolloid and the first solvation shell and therefore gives a first estimate for the hydrodynamic radius,
\begin{eqnarray}
R_H^\text{RDF} &=& r_m - R_S\\
&\approx& 2.7\,\sigma.
\end{eqnarray}
with $ R_S = 0.5\,\sigma $, the radius of the fluid particles. 

To get a more reliable value for the hydrodynamic radius, we use the finite size scaling of the diffusion constant in a system with periodic boundary conditions \cite{Dunweg1993}:
\begin{equation}
D_L = D_\infty  \left (1-2.837 \frac{R}{L_\text{B}}\right ),
\label{eq:fs_scaling}
\end{equation}
with the asymptotic diffusion coefficient $ D_\infty $ for an infinite system \gj{and the box size $ L_\text{B} $}. The value $ D_\infty $ can be determined from the zero-frequency limit of the memory function $ \hat{\gamma}(\omega) $ using the Einstein relation:
\begin{equation}
D_\infty = \frac{k_B T}{\gamma} = \frac{k_B T}{\gj{\hat{\gamma}}(\omega \rightarrow 0)} = \frac{k_B T}{6 \pi \eta R}.
\end{equation}

The finite-size diffusion constant $ D_L $ is calculated using a Green-Kubo relation, by integrating over the velocity autocorrelation function (VACF) of a solitary nanocolloid:
\begin{equation}
D_L = \frac{1}{3}\int_{0}^{\infty} \text{d}t' \left \langle \mathbf{v}(t')\mathbf{v}(0) \right \rangle,
\end{equation}
with $ \mathbf{v}(t) $ the velocity of the nanocolloid. 

\begin{figure}
\includegraphics[scale=1]{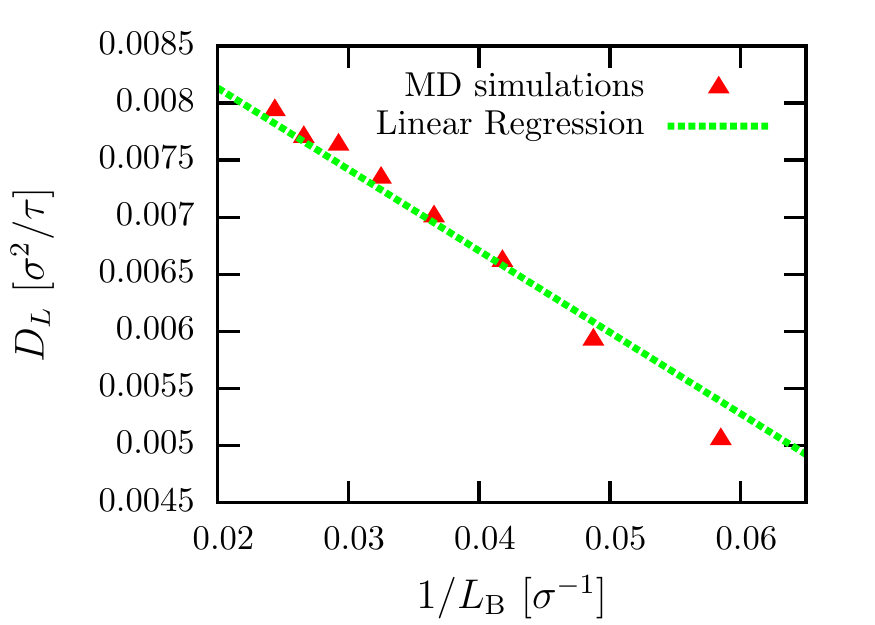}
\caption{Finite size scaling of the diffusion constant $ D_L $ of a single nanocolloid for different box sizes $ L_\text{B} $. The linear regression yields a hydrodynamic radius  $ R_H^D = (2.63 \pm 0.02)\,\sigma  $ by applying Eq.~(\ref{eq:fs_scaling}).}
\label{fig:diffusion}
\end{figure}

The resulting diffusion constants $ D_L $ for different inverse box sizes $ 1/L_\text{B} $ are presented in Fig.~\ref{fig:diffusion}. As expected, one can observe a linear behavior, allowing us to determine $ R_H^D $ by linear regression:
\begin{equation}
 R_H^D = (2.63 \pm 0.02)\,\sigma.
\end{equation}
In the following, we will use the value $  R_H =  R_H^D = 2.63~\sigma $ as reference value for the hydrodynamic radius.

\subsection{Speed of sound}

The adiabatic speed of sound $ c_0 $ is defined as,
\begin{equation}
c_0 = \sqrt{\left (\frac{\partial p}{\partial \rho}\right )_S} = \sqrt{\frac{1}{\rho \beta_S}} = \sqrt{\frac{C_P}{\rho \beta_T C_V}},
\end{equation}
with the isentropic compressibility $ \beta_S $, isothermal compressibility $ \beta_T $ \gj{and the heat capacities at constant pressure $ C_P $ and constant volume $ C_V $, respectively}. These thermodynamic observables can be determined by equilibrium fluctuations \cite{Allen1989}:
\begin{eqnarray}
\left\langle \delta \mathcal{H}^2 \right\rangle_{NVT} &=& k_B T^2 C_V,\\
\left\langle \delta V^2 \right\rangle_{NPT} &=& V k_B T \beta_T,\\
\left\langle \delta (\mathcal{H}+pV)^2 \right\rangle_{NPT} &=& k_B T^2 C_P,
\end{eqnarray}
with the total energy $ \mathcal{H} $. The subscripts connected to the equilibrium fluctuations denote the thermodynamic ensembles.

We therefore performed two MD simulations in the isothermal and isothermal-isobaric ensemble at box size $ L_\text{B}=43.09\,\sigma $ or $ \left\langle L_\text{B}\right\rangle =43.09 \,\sigma $, respectively, by applying a Nose-Hoover style thermo- and barostat \cite{Shinoda2004}. The results for the thermodynamic observables and the speed of sound can be found in Tab.~\ref{tab:sound_velocity}.

\begin{table}
\begin{tabular}{|c|c|c|c|}
\hline $ C_V/N $ & $ \beta_T $ $ \gj{[\sigma^3/\epsilon]} $ & $ C_P/N $ & $ c_0 $ $ \gj{[\sigma/\tau]} $ \\ 
\hline 2.38  & 0.076 &  $  4.59 $ & 5.63 \\
\hline 
\end{tabular} 
\caption{Several thermodynamic observables and speed of sound of a Lennard-Jones fluid with $ T^* = 1 $ and $ \rho^* = 0.8 $.}
\label{tab:sound_velocity}
\end{table}


\section{Generalized Langevin Equation and memory functions}
\label{sec:memory}

In Sec.~\ref{sec:theory} we already derived the memory function that determines the dissipative motion of a pair of spherical particles (see Eqs.~(\ref{eq:hydrotheory_first})-(\ref{eq:memory_theory})). This memory function can be used as input for the generalized Langevin equation (GLE),
\begin{equation}
\mathbf{F}(t) = M \mathbf{\dot{v}}(t) = \mathbf{F}_\text{C}(\mathbf{x}(t)) - \int_{-\infty}^{t} \text{d}t' \gamma(t-t') \mathbf{v}(t') + \partial \mathbf{F}(t),
\label{eq:GLE}
\end{equation}
with the conservative force $ \mathbf{F}_\text{C}(\mathbf{x}(t))  $, the memory function matrix $ \gamma(t) $ and the random force $ \partial\mathbf{F}(t) $, given by the fluctuation dissipation theorem (FDT),
\begin{equation}
\left\langle \partial\mathbf{F}(t) \partial\mathbf{F}(t') \right\rangle = k_\text{B} T \gamma(t-t').
\label{eq:FDT}
\end{equation}

The GLE is the outcome of the Mori-Zwanzig formalism, a theoretical tool to understand the process of coarse-graining a microscopic system \cite{Zwanzig1961,Mori1965,Zwanzig2001}. In our case, the systematic coarse-graining procedure connects a system with explicit solvent to an effective model that contains colloidal particles only (implicit solvent model). 

In this paper we utilize the GLE twofold: (a) We derive theoretical predictions for auto- and cross-correlation functions using the memory kernels derived in Sec.~\ref{sec:theory} and (b) we invert the derived equations to determine memory kernels from MD simulations. In the following analysis, we focus on the parallel dynamics, without explicitly including the subscript $ \parallel $. For the orthogonal component, the calculations are equivalent. For the conservative force we assume the particles to be trapped in a harmonic potential, $ \mathbf{F}_\text{C}(\mathbf{x}(t)) = - K (\mathbf{x}(t)-\mathbf{x}_0) = -K \delta \mathbf{x}(t)$, allowing us to assume a constant distance between the particles.

The starting point of the derivation are the equations of motion for the first particle,
\begin{eqnarray}
F_1(t) =  -K \delta x_1(t) &-& \int_{0}^{t} \text{d}s [ \gamma_{11}(t-s) v_1(s) \nonumber \\&+& \gamma_{12}(t-s) v_2(s) ] + \partial F_1(t).
\label{eq:GLE2}
\end{eqnarray}
This can be transformed into noise-free equations for the velocity auto-correlation function by multiplying with $ v_1(0) $ and taking the time-average. Using the relation $ \left\langle \partial \mathbf{F}(t) \mathbf{v}(0) \right\rangle = 0   $ we get,
\begin{eqnarray}
M \dot{C}^V_{11}(t) =  &-& \int_{0}^{t} \text{d}s [ K C^V_{11}(s) + \gamma_{11}(t-s) C^V_{11}(s) \nonumber \\&+& \gamma_{12}(t-s) C^V_{12}(s) ], \text{ for } t\geq0,
\label{eq:GLE3}
\end{eqnarray}
with $ C^V_{ij}(t) = \left\langle v_i(t) v_j(0) \right\rangle  $. Similarly, one can multiply Eq.~(\ref{eq:GLE2}) with $ v_2(0) $ to get the analogon of Eq.~(\ref{eq:GLE3}) for the cross-correlation function. These coupled differential equations can be decoupled by constructing equations for the relative and additive velocity correlation functions $ C^V_{\pm}(t) $, 
\begin{equation}
M \dot{C}^V_{\pm}(t) =  - \int_{0}^{t} \text{d}s [ K C^V_{\pm}(s) + \gamma_{\pm}(t-s) C^V_{\pm}(s)],
\label{eq:GLE4}
\end{equation}
with $ C^V_{\pm}(t) = C^V_{11}(t) \pm C^V_{12}(t) $ and $ \gamma_{\pm}(t) = \gamma_{11}(t) \pm \gamma_{12}(t) $. Starting from this equation, we can now derive relations to solve the previously mentioned tasks a) and b):
\begin{enumerate}[label=\alph*)]
\item By one-sided Fourier transform of Eq.~(\ref{eq:GLE4}) we can derive relations for auto- and cross-correlations in Fourier space (similar to Ref.~\cite{Theers2016}):
\begin{equation}
{\tilde{C}}^V_{\pm}(\omega) = \frac{k_B T}{-{\rm i}\omega M -\frac{K}{{\rm i}\omega}+\tilde{\gamma}_\pm(\omega)} .
\label{eq:FT_VACF}
\end{equation}
Here, $ \tilde{C}^V_{\pm}(\omega) $ denotes the one-sided Fourier transform of the velocity correlation functions,
\begin{equation}
\tilde{C}^V_{\pm}(\omega)  = \int_{0}^{\infty} \text{d}t e^{{\rm i} \omega t} C^V_{\pm}(t),
\end{equation}
similarly $ \tilde{\gamma}_\pm(\omega) $ the one-sided Fourier transform of the memory function and we have used $ M C^V_{\pm}(0) = M \left\langle v^2 \right\rangle = k_\text{B}T $. For $  C^V_{\pm}(-t) = C^V_{\pm}(t) = {C^V_{\pm}}^*(t)$, the one-sided Fourier transform can be related to the Fourier transform.
\begin{equation}
\hat{C}^V_{\pm}(\omega) = \int_{-\infty}^{\infty} \text{d}t e^{{\rm i} \omega t} C^V_{\pm}(t) = 2 \text{Re} \left\lbrace \tilde{C}^V_{\pm}(\omega) \right\rbrace 
\label{eq:FT}.
\end{equation}
To compare theory and simulations, we can therefore numerically invert the Fourier transform of the correlation functions in Eq.~(\ref{eq:FT_VACF}) and the Fourier transform of the FDT,
\begin{equation}
\tilde{C}^{\partial F}_{ij}(\omega) = k_B T \tilde{\gamma}_{ij}(\omega)  ,
\end{equation}
using Eq.~(\ref{eq:FT}).

\item Similar to Shin \emph{et al.} \cite{Shin2010} we take the derivative of Eq.~(\ref{eq:GLE4}) to write down a Volterra equation of second kind:
\begin{eqnarray}
M \ddot{C}^V_{\pm}(t) =  &-& K C^V_{\pm}(t) - \gamma_{\pm}(t) C^V_{\pm}(0) \nonumber\\&-& \int_{0}^{t} \text{d}s \gamma_{\pm}(t-s) \dot{C}^V_{\pm}(s).
\end{eqnarray}
This equation can be inverted numerically to determine the memory function by the recursive algorithm,
\begin{eqnarray}
\gamma_\pm(k\Delta t) &=& \left\lbrace C^V_{\pm}(0) + \frac{\Delta t}{2 M} C^{VF}_\pm(0) \right\rbrace ^{-1} \label{eq:volterra_algorithm}\\
 && \times \left \lbrace \frac{1}{M}C^{F}_\pm(k\Delta t) - K C^{V}_\pm(k\Delta t) \vphantom{\sum_{j=0}^{k-1}} \right.\ \nonumber \\ 
 && \left.\ - \Delta t \sum_{j=0}^{k-1} w_j C^{VF}_\pm((k-j)\Delta t) \gamma_\pm (j\Delta t) \right \rbrace ,\nonumber
\end{eqnarray}
where $ w_j = 0.5 $ for $ j=0 $ and $ w_j = 1 $ otherwise is the weight-factor for the numerical integration. We also introduced the relative and additive force and velocity-force correlation functions, 
\begin{eqnarray}
 C^{F}_\pm(t) &=& \left\langle F_1(t) F_1(0) \right\rangle  \pm \left\langle F_1(t) F_2(0) \right\rangle \nonumber \\ &=&  - M ^2 \ddot{C}^V_{\pm}(t)  \\  C^{VF}_\pm(t) &=& \left\langle F_1(t) v_1(0) \right\rangle  \pm \left\langle F_1(t) v_2(0) \right\rangle \nonumber \\ &=& M \dot{C}^V_{\pm}(t).
\end{eqnarray}
 The initial condition for the above algorithm is given by,
\begin{equation}
\gamma_\pm(0) = \frac{1}{M} \frac{C^{F}_\pm(0)}{C^{V}_\pm(0)} - K.
\end{equation}
\end{enumerate}

\section{Results}
\label{sec:results}

In the following, the results of the theoretical evaluation will be presented and compared to numerical experiments. First, we will analyze the differences between the two theoretical approaches (point force and unsteady velocity field) derived in Sec.~\ref{sec:theory}. Then we will compare the theoretical results to MD simulations of nanocolloids in a Lennard-Jones fluid. 

\subsection{Hydrodynamic theory}

\begin{figure}
\includegraphics[scale=1]{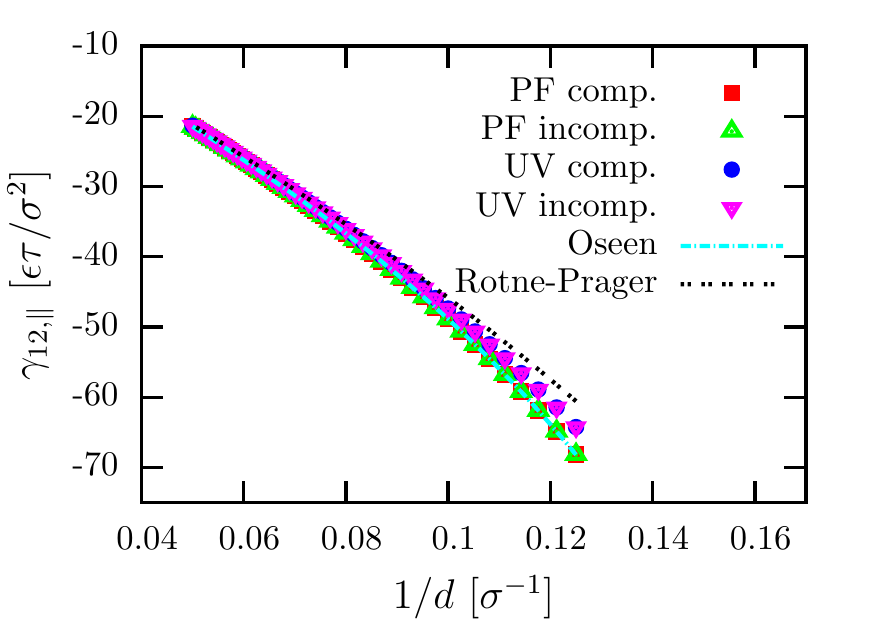}
\caption{Distance-dependence of the integrated memory kernels using the point force (PF) and the unsteady velocity field (UV) approach. The results from hydrodynamic theory presented in this paper are compared to friction tensors derived from the well-known Oseen and Rotne-Prager tensors \cite{Oseen1911,Rotne1969}. The incompressible solution corresponds to the limit $ c_0 \rightarrow \infty $.}
\label{fig:comparison_methods_distance}
\end{figure}

Before analyzing the time-dependence of the memory function, we compare the results presented in this paper to the well-known distance-dependent hydrodynamic interaction tensors: the Oseen tensor,
\begin{eqnarray}
D_{ii}^\text{Oseen}(d) &=& - \frac{1}{6 \pi \eta R},\\
D_{ij}^\text{Oseen}(d) &=&  \frac{1}{4 \pi \eta R} \left(  \frac{R}{d}\right) , i \neq j,
\end{eqnarray}
and the Rotne-Prager (RP) tensor,
\begin{eqnarray}
D_{ii}^\text{RP}(d) &=&  - \frac{1}{6 \pi \eta R},\\
D_{ij}^\text{RP}(d) &=&  \frac{1}{4 \pi \eta R} \left(  \frac{R}{d} - \frac{2}{3} \left( \frac{R}{d}\right)^3 \right) , i \neq j,
\end{eqnarray}
\gj{with the distance  $ d $ between the centers of the nanocolloids of radius $ R $.}
For the comparison, we have to invert these diffusion tensors to derive the hydrodynamic \gj{friction} tensors,
\begin{equation}
F_{i,\parallel} = -\gamma_{ij} v_{j,\parallel},
\end{equation}
with
\begin{eqnarray}
\gamma_{ii}^\text{Oseen}(d) &=&  \frac{24 \pi R \eta}{4  - 9 ( \frac{R}{d}) ^2},\\
\gamma_{ij}^\text{Oseen}(d) &=& - \frac{1}{d} \frac{36 \pi R^2 \eta}{4  - 9 ( \frac{R}{d}) ^2}, i \neq j,
\end{eqnarray}
and 
\begin{eqnarray}
\gamma_{ii}^\text{RP}(d) &=&  \frac{24 \pi R \eta}{4  - 9(\frac{R}{d}) ^2 + 12( \frac{R}{d}) ^4 - 4( \frac{R}{d}) ^6 },\\
\gamma_{ij}^\text{RP}(d) &=&  -\frac{1}{d} \frac{36 \pi R^2 \eta (1-\frac{2}{3}( \frac{R}{d}) ^2)}{4  - 9 ( \frac{R}{d}) ^2 + 12( \frac{R}{d}) ^4 - 4 (\frac{R}{d}) ^6}, i\neq j.
\end{eqnarray}
A comparison between the off-diagonal component of $ \gamma_{ij}^\text{Oseen} $ and $ \gamma_{ij}^\text{RP} $ to the time-integrated memory kernel,
\begin{equation}
\gamma_{12,\parallel} = \int_{0}^{\infty} \text{d} t \gamma_{12,\parallel}(t) = \hat{\gamma}_{12,\parallel}(\omega=0),
\end{equation}
can be found in Fig.~\ref{fig:comparison_methods_distance}. The figure shows that the point force approach and the Oseen friction tensor are equivalent when applying the Markovian approximation and integrating out the time-dependence. Furthermore, the unsteady velocity field approach already represents a significant improvement over the point force method but is not yet similar to the Rotne-Prager approximation. The improvement can be explained by the use of the correct boundary conditions for sphere 1 (see Eq.~(\ref{eq:boundary_uvf})). The flow-field is, however, evaluated at the center of sphere 2, which does not take the correct boundary conditions for sphere 2 into account (see Eqs.~(\ref{eq:uvf_result_par})-(\ref{eq:uvf_result_par_l})). This explains the deviation from the Rotne-Prager solution.

\begin{figure}
\includegraphics[scale=1]{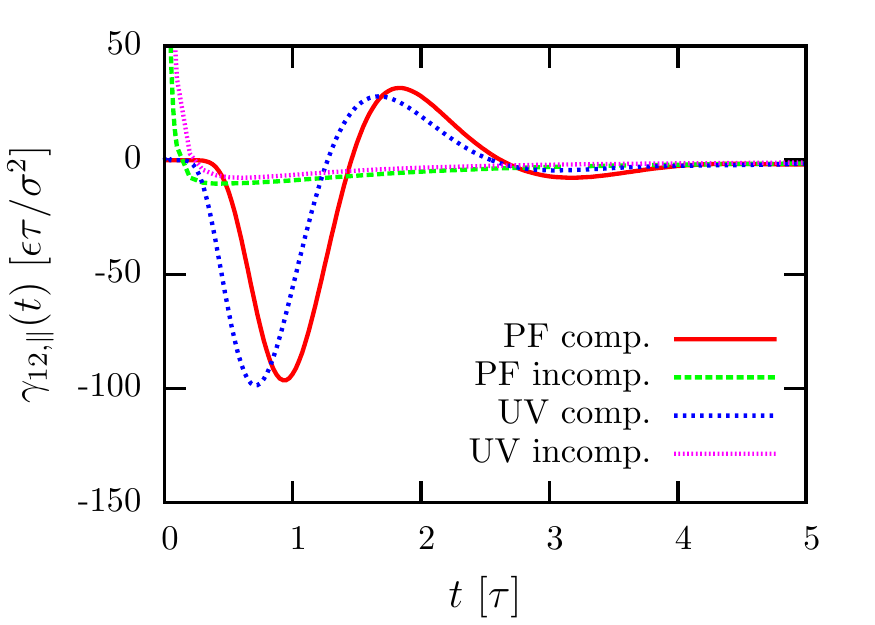}
\caption{Inverse Fourier transform of the frequency-dependent interactions derived in Sec.~\ref{sec:theory}. The distance between the two spheres was chosen to be $ d=8.5\,\sigma $.}
\label{fig:comparison_methods}
\end{figure}

Fig.~\ref{fig:comparison_methods} shows the inverse Fourier transform of the theoretically derived frequency-dependent hydrodynamic interactions. An unphysical instantaneous interaction between the two spheres can be observed in the incompressible limit ($ c_0\rightarrow \infty $). In contrast, for a more realistic compressible fluid, the interaction between the two spheres is mediated by a sound wave with finite speed of sound $ c_0 $. The huge difference between the two limits shows, that the compressibility has an important impact on the time-dependence of the interaction. In fact, the interaction through sound waves seems to dominate, at least for small distances. This is a noteworthy result, because the frequency-dependent self-interaction of a solitary sphere can be very well described by an incompressible fluid. To summarize, we conclude that the self-interaction is dominated by transversal waves, while the pair-interaction is dominated by longitudinal waves, \gj{at least for the case of nanocolloids}.

\subsection{Comparison to MD simulations}

The results discussed in the last section can now be compared to MD simulations. To this end, we performed large-scale simulations of the system described in Sec.~\ref{sec:fluid} with box size $ L_B = 51.2993\,\sigma $ and harmonic constant $ K = 5 \epsilon \sigma^{-2} $. We applied the algorithm derived in Eq.~(\ref{eq:volterra_algorithm}) to reconstruct the memory kernel from the MD results and determined the velocity cross-correlation function $ C^V_{12,\parallel}(t) $. Additionally, we show results for the distance-dependence of the memory self-correlation function.

\subsubsection{Cross-Correlations}

\begin{figure}
\includegraphics[scale=1]{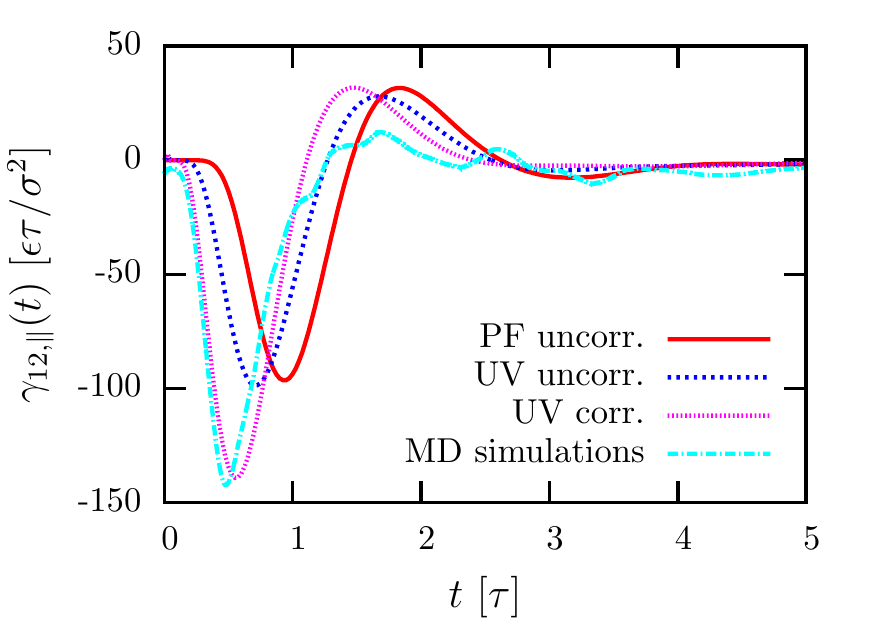}
\caption{Comparison between the results from hydrodynamic theory and the memory functions obtained from MD simulations. The distance between the two spheres was chosen to be $ d=8.5\,\sigma $. The corrected theory includes a distance-shift of $ \Delta d = 1.0\,\sigma $.}
\label{fig:comparison_theory_md}
\end{figure}
The results presented in Fig.~\ref{fig:comparison_theory_md} illustrate that the correspondence between theory and simulations is not yet perfect. The shock wave induced by the first colloid hits the second colloid at smaller times than expected by theory. This can be explained by the already mentioned discrepancies in the definition of the boundary conditions. To correct for this problem, we introduced a small distance-shift of $ \Delta d = 1.0\,\sigma $ to the solution of the unsteady velocity field. While this does not completely resolve the differences, it shows that the theory is able to reproduce the results from MD simulations fairly accurately. 

\begin{figure}
\includegraphics[scale=1]{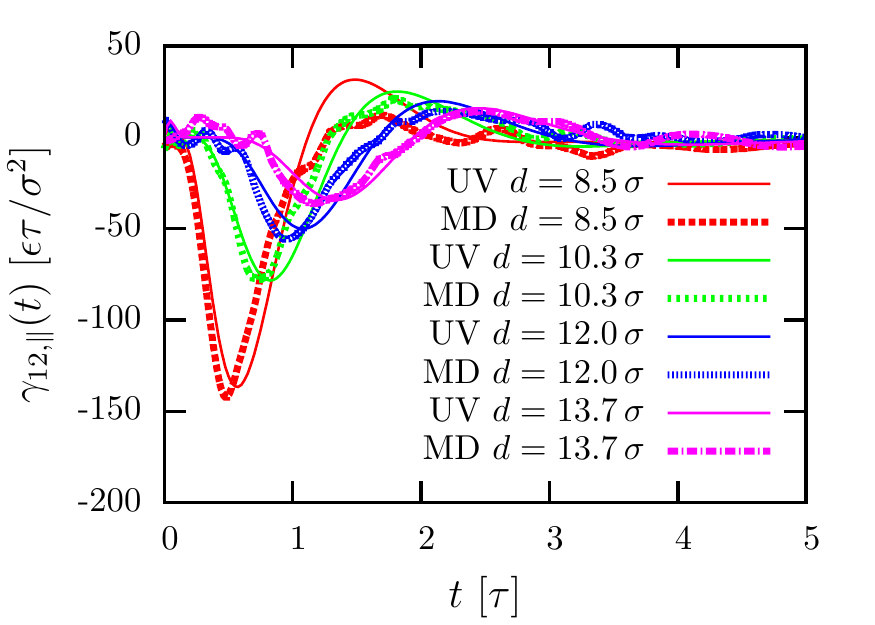}
\includegraphics[scale=1]{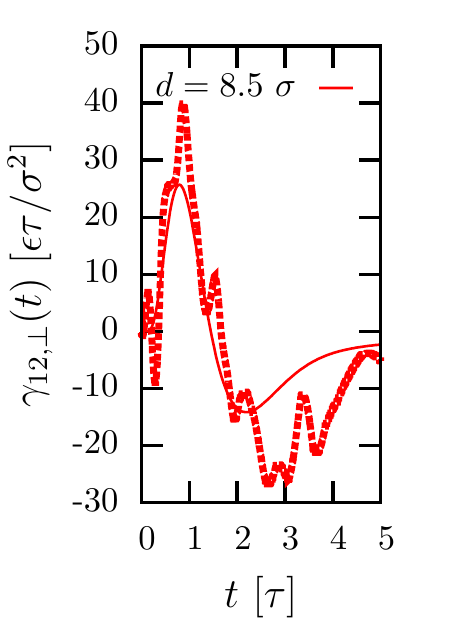} \includegraphics[scale=1]{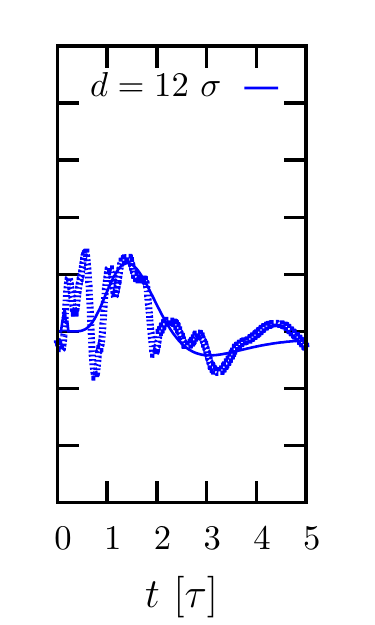}
\caption{Distance-dependence of the cross-memory function. Similar to Fig.~\ref{fig:comparison_theory_md} a distance-shift of $ \Delta d = 1.0\,\sigma $ was included. The upper panel shows the parallel component $ (\parallel) $ and the lower panels the orthogonal component $ (\perp) $. }
\label{fig:distance_dependence_memory}
\end{figure}

The discrepancies between theory and simulations almost vanish at larger particle distances (see Fig.~\ref{fig:distance_dependence_memory}, upper panel). Especially for $ d > 4 R $ the agreement is remarkable. This observation therefore confirms the statements made in the previous paragraph and shows that we are indeed able to precisely model the longitudinal waves that mediate the interaction between the two nanocolloids. This statement also holds for the orthogonal component of the cross-memory function (see Fig.~\ref{fig:distance_dependence_memory}, lower panels). The statistical errors of the data are very large due to the small amplitude of the cross-correlations. Nevertheless, the theoretical curve and simulation results agree very well. 

It is important to note that the results from MD simulations are affected by finite size effects. The impact of these are, however, not distinguishable from the statistical noise due to sampling accuracy.
\begin{figure}
\includegraphics[scale=1]{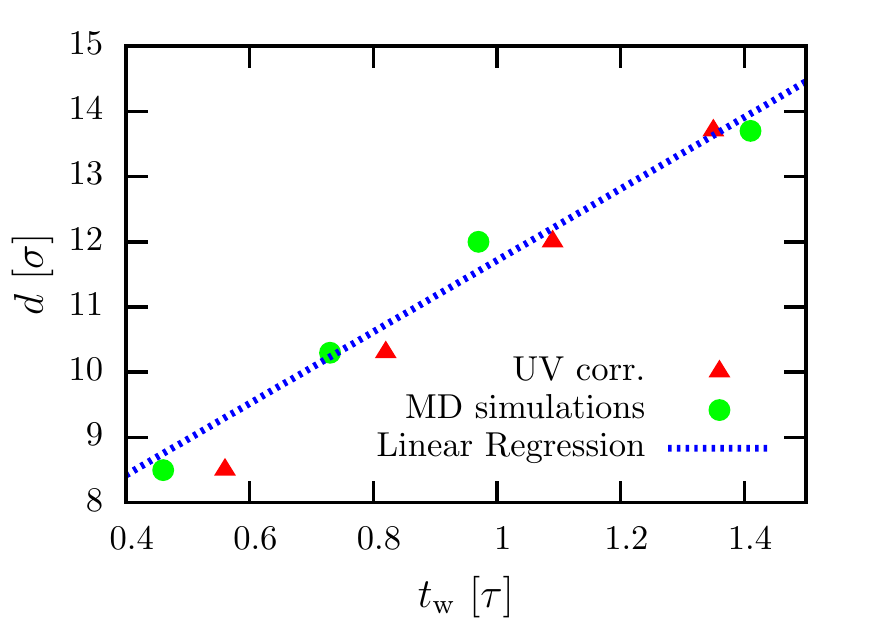}
\caption{ Distance-dependence of the time $ t_w $ needed by the sound wave to propagate through the medium and interact with another nanocolloid. As expected, the slope of the linear curve is similar to the speed of sound $ c_0^\text{fit} = 5.48 $ and the intersection with the y-axis corresponds to the diameter of the nanocolloids $ 2 R_H^\text{fit} = 6.23 $.   }
\label{fig:sound_wave}
\end{figure}

The observation that the interaction is dominated by a sound wave can be used to determine the speed of sound $ c_0 $ in a straightforward analysis. In Fig.\,\ref{fig:sound_wave} the distance-dependence of the time $ t_w $ needed by the sound wave to mediate the interaction is illustrated. This time is defined as the minimum of the cross memory function $ \gamma_{12,\parallel}(t) $. The figure shows a linear dependence and therefore allows for a calculation of the speed of sound by linear regression. Additionally, one can get an estimate of the hydrodynamic radius from the intersection of the line with the y-axis.
\begin{figure}
\includegraphics[scale=1]{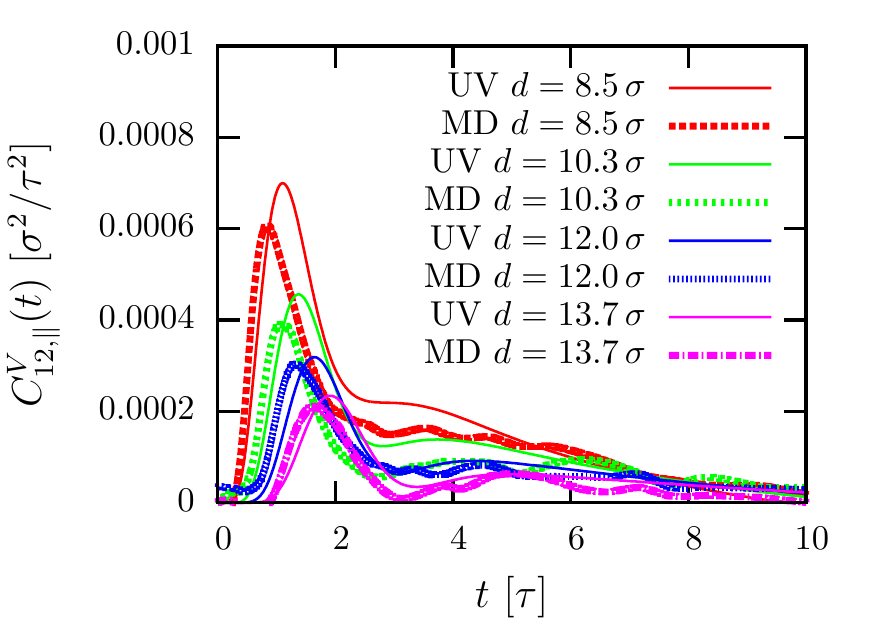}
\caption{Distance-dependence of the velocity cross-correlation function. Similar to Fig.~\ref{fig:comparison_theory_md} a distance-shift of $ \Delta d = 1.0\,\sigma $ was included. }
\label{fig:distance_dependence_vacf}
\end{figure}

Fig.\,\ref{fig:distance_dependence_vacf} shows the distance-dependence of the velocity cross-correlation function. While the qualitative agreement between theory and simulations is very good, there are larger deviations for medium times. This can be explained by the fact that the frequency-dependent response of the solitary sphere $ \hat{\gamma}(\omega) $ enters the velocity cross-correlation directly (see Eq.\,(\ref{eq:FT_VACF})). For high frequencies, this will therefore lead to significant deviations from simulations. As mentioned above, this will not affect the memory kernel, since the multiplication with the interaction parameter $ D_{\parallel/\perp}(\omega) $ acts as low-pass filter (see Sec.~\ref{sec:theoryUV} and Fig.~\ref{fig:frequency_dependence}).

\subsubsection{Auto-Correlations}

In this section we analyze whether the self-memory kernel also shows a dependence on the distance of the two nanocolloids.  For an extensive analysis of the auto-correlations of a solitary colloid, we refer to Ref.~\cite{Theers2016}.
\begin{figure}
\includegraphics[scale=1]{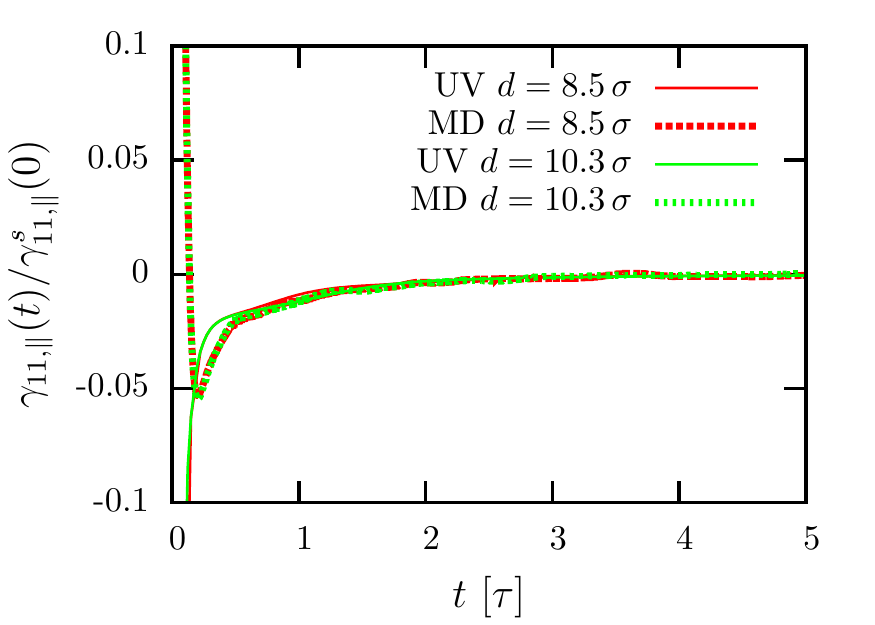}
\includegraphics[scale=1]{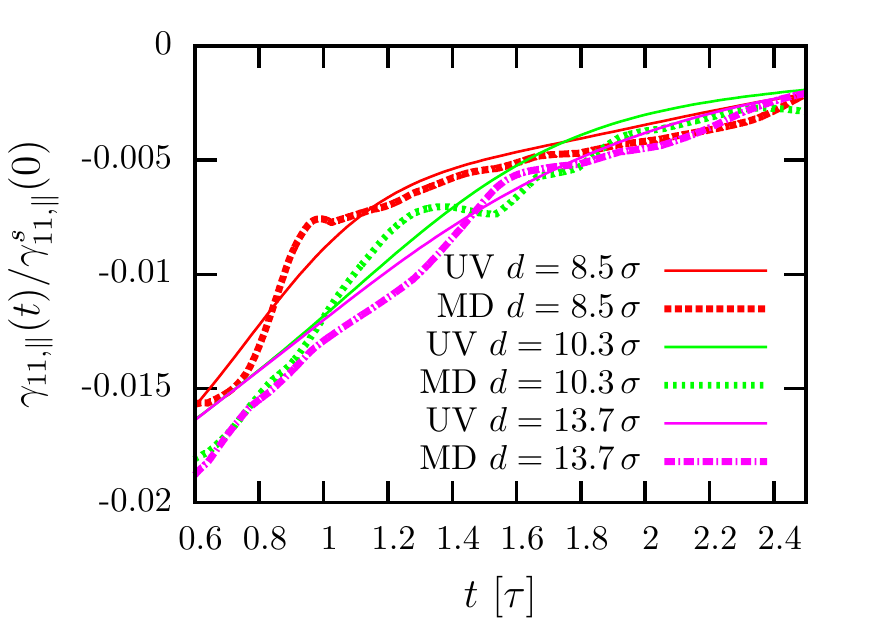}
\caption{Distance-dependence of the normalized self-memory kernel. Similar to Fig.~\ref{fig:comparison_theory_md} a distance-shift of $ \Delta d = 1.0\,\sigma $ was included. All curves are normalized by the simulation value $ \gamma^s_{11,\parallel}(t=0) $. The upper and lower figure show the same data with different zoom. }
\label{fig:self_memory}
\end{figure}
The upper panel of Fig.~\ref{fig:self_memory} shows the normalized self-memory kernel in coarse resolution. This figure illustrates that there is no significant difference between the self-memory kernels for different particle distances. When comparing theory and simulations one observes completely different behavior for small times and good agreement for larger times. The discrepancy at small times is related to the already mentioned difference between particle and continuum description. The long time hydrodynamic properties are, however, perfectly reproduced.

The zoomed view of Fig.~\ref{fig:self_memory} (see bottom panel) demonstrates that there are indeed very small differences between the auto-correlations in the vicinity of other nanocolloids. These differences can be described with the hydrodynamic theory derived in this paper. In this fine resolution the statistical error of the simulations are visible, however, it is still possible to distinguish the curve for the smallest distance $ d= 8.5\,\sigma $ from the curves for $ d > 10\,\sigma $. 


\subsection{Identification of units and time scales}

\label{sec:scales}

\gj{To interpret our results, we will first qualitatively map the LJ fluid to water by identifying the units of length $ \sigma $, energy $ \epsilon $ and time $ \tau $, respectively. This enables us to determine the LJ fluid transport coefficients. In a second step we compare the time scales of our simulations with realistic systems of colloids and nanocolloids.}

\subsubsection{Mapping of reduced units}

\gj{To perform a mapping of the reduced system to water we use three assumptions:
\begin{itemize}
\item the mass $ m $ of a LJ particle corresponds to $ 18\,u $ (the mass of a water molecule)
\item the mass density of the LJ fluid corresponds to $ 1000\,\text{kg}/\text{m}^3 $ (the density of water)
\item the temperature $ T = \epsilon/k_\text{B} $ corresponds to $ 300\,\text{K} $.
\end{itemize}
\begin{table}
\begin{tabular}{|c|c|c|}
\hline   & this paper & Dysthe \emph{et al.}\cite{Dysthe2002}  \\ 
\hline $ \sigma $  & $ 0.29\cdot 10^{-9}\,\text{m} $ &  $ 0.29\cdot 10^{-9}\,\text{m} $  \\
\hline $ \epsilon $  & $ 4.0 \cdot 10^{-21}\,\text{J} $ &  $ 5.5 \cdot 10^{-21}\,\text{J} $  \\
\hline $ \tau $  & $ 8.0 \cdot 10^{-13}\,s $& -   \\
\hline 
\end{tabular} 
\caption{Mapping of the reduced units describing the LJ fluid studied in this paper to water. The results are compared to the triple point mapping performed in Ref.\,\cite{Dysthe2002}.}
\label{tab:mapping}
\end{table}
The results of this mapping can be found in Tab.\,\ref{tab:mapping}. The values correspond very precisely to results of Dysthe \emph{et al.}\cite{Dysthe2002}  achieved by mapping the triple points of both fluids.
\begin{table}
\begin{tabular}{|c|c|c|}
\hline   & LJ fluid & water ($ 30\,^{\circ}\text{C} $)  \\ 
\hline shear viscosity $ \eta $ $ [\epsilon \tau/\sigma^3] $  & $ 2.11 $ &  $ 6.23 $  \\
\hline bulk viscosity $ \zeta $ $ [\epsilon \tau/\sigma^3] $  & $ 0.88 $ &  $ 18.69 $  \\
\hline speed of sound $ c_0 $ $ [\sigma/\tau] $  & $ 5.63 $& $ 4.05 $   \\
\hline 
\end{tabular} 
\caption{Transport coefficients and speed of sound of the LJ fluid compared to water at room temperature. The mapping of units was performed according to Tab.\,\ref{tab:mapping}.}
\label{tab:real_parameter}
\end{table}
The conversion from reduced to real units allows us to compare the dynamical fluid properties (see Tab.\,\ref{tab:real_parameter}). These results indicate, that the model LJ fluid is similar to water. The only significant difference is the bulk viscosity $ \zeta $ of the LJ fluid, which is about 20 times smaller than in water. These differences will lead to quantitatively different results since longitudinal waves will decay faster, but the overall picture remains.}

\subsubsection{Comparison of the simulations with colloidal time scales}

\gj{In colloidal suspensions we can identify four different important time scales \cite{Padding2006}.
\begin{itemize}
\item The \emph{sonic time} $ \tau_s $ over which sound propagates one colloidal radius. This time scale describes the interaction of colloids by longitudinal waves.
\item The \emph{kinematic time} $ \tau_v $ over which momentum diffuses one colloidal radius. On this time scale, transversal waves propagate between the colloids.
\item The \emph{Brownian relaxation time} $ \tau_B $ over which the velocity correlation function of a colloid decays.
\item The \emph{colloid diffusion time} $ \tau_D $ over which a colloid diffuses over its radius.
\end{itemize} 
With the above determined fluid parameter, we can estimate these time scales for both nanocolloids ($ R\approx 1\,\text{nm} $) and colloids ($ R\approx 100\,\text{nm} $). In our simulations, the nanocolloids have a radius of $ R\approx1\,\text{nm} $ according to the mapping discussed above.}

\begin{table}
\begin{tabular}{|c|c|c|}
\hline    & colloid& nanocolloid  \\ 
\hline Sonic time $ \tau_s = R/c_0 $  & $ 10^{-10}\,s $& $ 10^{-12}\,s $   \\
\hline Kinematic time $ \tau_v = \rho R^2/\eta $  & $ 10^{-8}\,s $& $ 10^{-12}\,s $   \\
\hline Brownian relaxation time $ \tau_B = \rho R^2/\eta $  & $ 10^{-8}\,s $& $ 10^{-12}\,s $   \\
\hline Colloid diffusion time $ \tau_d = \eta R^3/k_\text{B} T $  & $ 10^{-3}\,s $& $ 10^{-9}\,s $   \\
\hline 
\end{tabular} 
\caption{Different time scales that are relevant in colloidal suspensions. The time scales of colloids ($ R\approx 100\,\text{nm} $) are compared to nanocolloids ($ R\approx 1\,\text{nm} $). }
\label{tab:time_scales}
\end{table}

\gj{The results are summarized in Tab.\,\ref{tab:time_scales}. For colloids there is a distinct time scale separation between the diffusion of the colloid and the Brownian relaxation time. This fact is often used when performing Brownian dynamics simulations (overdamped dynamics). Furthermore, the sonic time scale seems to be much smaller than the Brownian relaxation time, indicating, that the effects of compressibility may not be relevant on this time scale. However, this picture totally changes for nanocolloids. In this case, there is an overlapping of the sonic, the kinematic and the Brownian relaxation time. }

\gj{We can therefore conclude, that the consequences of compressibility on the hydrodynamic pair interaction are significant for nanocolloids, while for colloids differences will only be observable for very high frequencies.}


\section{Conclusions and Outlook}
\label{sec:conclusion}

In this paper we derived theoretical expressions for the hydrodynamic interaction of two solid spheres in a compressible fluid. We could show that the compressibility of the fluid has an important impact on the frequency-dependence of the hydrodynamic interaction. In fact, the most pronounced feature of the pair memory function that describes this hydrodynamic interaction is a sound wave that propagates with speed of sound $ c_0 $. Moreover, we determined pair memory functions from molecular dynamics simulations of two nanocolloids in a Lennard-Jones fluid and showed that there is very good agreement between simulations and theory. 

Our work can have an impact in various different ways. The results from hydrodynamic theory can be used as input to perform dynamically consistent coarse-grained simulations, for example in combination with the generalized Langevin equation \cite{Li2017,Jung2017}. Additionally, when it is experimentally possible to measure velocity correlation functions with optical tweezers \cite{Li2010}, the theory might be used to understand and evaluate experimental results. This could be seen as update to the classical two-point microrheology that targets the determination of fluid rheology by investigating the correlations of macromolecules submerged in this fluid.

\section*{Acknowledgment}

\fs{The authors want to thank Jay D. Schieber for helpful discussions concerning Ref. \cite{Cordoba2013}}. This work was funded by the German Science Foundation within project A3 of the SFB TRR 146. Computations were carried out on the Mogon Computing Cluster at ZDV Mainz.


\bibliography{library}

\end{document}